\documentclass{jpp}
\usepackage{graphicx}
\usepackage{hyperref}
\usepackage[normalem]{ulem}
\usepackage[utf8]{inputenc}
\usepackage[T1]{fontenc}
\usepackage{amsmath}
\usepackage{comment}
\usepackage{color}

\newcommand{\bd}{\boldsymbol}

\newcommand{\mc}{\mathcal}
\newcommand{\avg}[1]{\left\langle{#1}\right\rangle}
\newcommand{\pd}{\partial}
\newcommand{\etabar}{\bar{\eta}}
\newcommand{\rmc}{{\rm c}}
\newcommand{\rmb}{{\rm b}}
\newcommand{\rmd}{{\rm d}}
\newcommand{\rme}{{\rm e}}
\newcommand{\rmi}{{\rm i}}

\newcommand{\rms}{{\rm s}}
\newcommand{\rmt}{{\rm t}}

\shorttitle{Theory of Equivalent Tokamaks to Quasi-symmetric Stellarators}
\shortauthor{H. Zhu, R. Gaur, X. Wei, Z. Lin, and A. Bhattacharjee}

\title{Theory of Equivalent Tokamaks for Characterizing Turbulent Transport in Quasi-symmetric Stellarators}

\author{Hongxuan Zhu\aff{1}
  \corresp{\email{hongxuan@zju.edu.cn}},
  R. Gaur\aff{2}, X. Wei\aff{3}, Z. Lin\aff{4},
 \and A. Bhattacharjee\aff{5}}

\affiliation{\aff{1}School of Physics, Zhejiang University, Hangzhou, Zhejiang, China 310027
\aff{2}Department of Physics, University of Wisconsin, Madison, Wisconsin, USA 53706
\aff{3}Honghu Future Energy (Shanghai) Ltd., Shanghai, China  201108
\aff{4}Galaxy Energy LLC, Irvine, CA, USA 92612
\aff{5}Department of Astrophysical Sciences, Princeton University, Princeton, NJ, USA 08540}

\begin{document}

\maketitle

\begin{abstract}
It is well known that quasi-symmetric (QS) stellarators are isomorphic to tokamaks in terms of their neoclassical-transport properties, and the corresponding transport coefficients can be calculated in the same manner as in tokamaks. However, less is known regarding the turbulent-transport properties of QS stellarators, e.g. the transport coefficients from the ion-temperature-gradient (ITG) mode. In this work, a systematic theory of the ``equivalent tokamaks'' for QS stellarators is presented based on the local gyrokinetic formulation and the near-axis expansion  theory. It is shown that to zeroth order in the minor radius, the equivalent tokamaks can be chosen to have circular flux surfaces and can be characterized by three geometric quantities: the aspect ratio, the rotational transform, and the magnetic shear. To achieve first-order accuracy, however, not all QS stellarators have equivalent tokamaks, but good approximations can be found for some cases  either as global or local equilibria. Local and global gyrokinetic simulations of ITG transport are performed for a selection of QS configurations, and quantitative agreement in the turbulent transport levels is found between the stellarators and their equivalent tokamaks.
\end{abstract}

\section{Introduction}
Quasisymmetric (QS) stellarators are a type of magnetic confinement fusion device where the magnitude of the magnetic field vector $\bd{B}$, which lies on flux surfaces, can be expressed as $|\bd{B}|=B(\psi,M\theta-N\varphi)$ \citep{Boozer83,Nuhrenberg88,Rodriguez20}.  Here, $\psi$ is
the toroidal magnetic flux divided by $2\pi$, $\theta$ and $\varphi$ are the poloidal and toroidal angles in Boozer coordinates \citep{Boozer82}, and $M$ and $N$ are constant integers. This includes both quasi-axisymmetric (QA) devices where $M\neq 0$ and $N=0$, and quasi-helically (QH) symmetric devices where $M\neq 0$ and $N\neq 0$. (The quasi-poloidally symmetric devices with $M=0$ cannot be accurately achieved near the magnetic axis, and hence are not considered in this paper.) It is well known that many neoclassical-transport properties in QS stellarators can be calculated  from the formulation derived for tokamaks, which is due to the fact that the drift-kinetic guiding-center motion in QS stellarators is isomorphic to that in tokamaks in Boozer coordinates \citep{Boozer83}. This fact has been utilized for the design of QS stellarators. For example, \cite{LBD22} obtained numerical QS equilibria with self-consistent bootstrap currents, which were calculated from the Redl formula initially obtained for tokamak equilibria \citep{Redl21}.

While the neoclassical-transport theory has been well developed, less is known regarding the turbulent-transport properties of QS stellarators.  As the neoclassical transport in QS stellarators can be reduced to a low level similar to tokamaks, turbulent transport is expected to dominate. Since QS stellarators are very similar to tokamaks, the dominant types of drift-wave turbulence in tokamaks, \eg the ion-temperature-gradient (ITG) mode and the trapped-electron mode (TEM), are also present in QS stellarators \citep{Huang22}. Turbulent transport is highly sensitive to the geometric and plasma parameters. For tokamaks,  quasilinear transport models  such as \textsf{TGLF} \citep{Staebler07} exist, but direct gyrokinetic simulations often provide more accurate predictions. On the other hand, QS stellarators have a vast design parameter space and their transport level varies significantly.   Recently, through machine learning, \cite{Landreman25} identified the two most important geometric factors that correlate with the ITG transport across a large number of configurations: the flux-surface compression in the bad-curvature region and the geodesic curvature. Therefore, the development of reduced transport models for QS stellarators based on these theoretical insights will be highly desirable.

In order to develop reduced models for the turbulent transport in QS stellarators, one may borrow the experience from studying neoclassical transport and make an analogy with tokamaks, which are simpler in their geometry and hence can be characterized using  fewer parameters compared to stellarators. This idea has been previously explored by \cite{Nakata20} who compared the ITG turbulent transport in the CFQS stellarator with an ``equivalent tokamak'', but has not been fully developed into a systematic theory. Nevertheless, previous studies have suggested that the transport properties of QS stellarators can indeed be characterized by only a few parameters. For example, by training an autoencoder, \cite{Wei26} found that a large database of QH stellarators can be approximately represented by as few as only 3 parameters. The Rosenbluth--Hinton (RH) residual flow level \citep{Rosenbluth98}, which is related to the zonal-flow strength, has also been derived analytically for QS stellarators and depends essentially on only two parameters: the rotational transform $\iota$ and the helicity number $N$ \citep{Plunk24,Rodriguez25RH,Zhu25}. The RH residual has been found to be strongly correlated with the transport level \citep{Chen25,Wei26}. In particular, the vast difference in the RH residual between QA and QH configurations may help explain the difference in their ITG transport levels \citep{Mckinney19,Chen25}. We note that even though the neoclassical-transport coefficients can be calculated from the formula derived for tokamaks, the question remains unanswered whether such tokamaks can be achieved as local or global 3-dimensional magnetohydrodynamic (MHD) equilibria.

In this work, a systematic theory of the equivalent tokamaks for QS stellarators is presented based on the local gyrokinetic formulation and the near-axis-expansion (NAE) theory. For ideally QS stellarators, geometric quantities such as the magnetic-field strength and the magnetic-drift frequency are independent of  the field-line label $\alpha=\theta-\iota\zeta$, where $\iota$ is the rotational transform and $\theta$ and $\zeta$ are the straight-field-line poloidal and toroidal angles, respectively \citep{NiesProposal}. The $\alpha$-dependent geometric quantities only appear due to the finite-Larmor-radius (FLR) effects. Therefore, the equivalent tokamak is defined as an axisymmetric MHD equilibrium where the $\alpha$-independent quantities are identical to those of the QS stellarator. It is shown that to zeroth-order accuracy in the minor radius the equivalent tokamaks can be chosen to have circular flux surfaces and can be characterized by three geometric quantities: the aspect ratio $R/a$, the rotational transform $\iota$, and the magnetic shear $\hat{s}$. To achieve first-order accuracy, however, not all QS stellarators have equivalent tokamaks, but good approximations can be found for some cases either as global equilibria, or as local equilibria using, e.g., the Miller model \citep{Miller98}. Finally, simulations of ITG turbulent transport are performed for a selection of QS configurations using the local gyrokinetic code \textsf{GX} \citep{Mandell24} and global gyrokinetic code \textsf{GTC}\footnote{\url{https://sun.ps.uci.edu/gtc}}, which demonstrate quantitative agreement between the stellarators and their equivalent tokamaks.

The rest of the paper is organized as follows. Theoretical background on the local gyrokinetic equations,  the NAE theory, and the Miller model are presented in section~\ref{sec:background}. The derivation of equivalent tokamaks from the NAE is presented in section~\ref{sec:theory}. Applications of the equivalent-tokamak theory to several representative QS configurations are given in section~\ref{sec:application}. Numerical results from gyrokinetic simulations are shown in section~\ref{sec:numerical}. Conclusions are given in section~\ref{sec:conclusion}. Details on the derivation of the coefficient $B_{20}$ from the NAE theory are presented in Appendix~\ref{app:B20}.
\section{Theoretical background}
\label{sec:background}
\subsection{The gyrokinetic equation}
The gyrokinetic equation describes the evolution of the 5-dimensional distribution function of gyro-centers in a magnetic field $\bd{B}$, which can be represented as
\begin{equation}
\label{eq:theory_covariant_B}
\bd{B}=\nabla\psi\times\nabla\alpha=G(\psi)\nabla\zeta+I(\psi)\nabla\theta+\delta(\psi,\theta,\zeta)\nabla\psi.
\end{equation}
Here, $\psi$ is the toroidal magnetic flux divided by $2\pi$, $\alpha=\theta-\iota\zeta$ is the field-line label, and $\iota(\psi)$ is the rotational transform. Consider the species $s$ with mass $m_s$ and charge number $Z_s$ and use $(\bd{R},\varepsilon,\mu)$ as the phase-space variables where $\bd{R}$ is the gyro-center position, $\varepsilon$ is the energy, and $\mu$ is the magnetic moment. The collisionless, electrostatic, nonlinear gyrokinetic equation  can be written as \citep{Highcock12,Abel13}
\begin{equation}
    \frac{\pd h_s}{\pd t}+\left(v_\parallel\hat{\bd{b}}+\bd{v}_D+\frac{1}{B}\hat{\bd{b}}\times\nabla\avg{\varphi}_{\bd{R}}\right)\cdot\nabla h_s=\frac{Z_seF_s}{T_s}\frac{\pd\avg{\varphi}_{\bd{R}}}{\pd t}-\frac{\partial F_s}{\partial\psi}\frac{1}{B}\hat{\bd{b}}\times\nabla\avg{\varphi}_{\bd{R}}\cdot\nabla\psi.
\end{equation}
Here, $h_s(t,\bd{R},\varepsilon,\mu)$ is the non-adiabatic part of the perturbation in the gyro-center distribution function, $v_\parallel=\pm\sqrt{2(\varepsilon-\mu B)/m_s}$ is the parallel velocity,  $\hat{\bd{b}}=\bd{B}/B$,
\begin{equation}
    \bd{v}_D=\frac{\hat{\bd{b}}}{\Omega_s}\times\left(v_\parallel^2\bd{\kappa}+\frac{1}{2}v_\perp^2\nabla\ln B\right),
\end{equation}
is the magnetic-drift velocity.  Here, $\bd{\kappa}=\hat{\bd{b}}\cdot\nabla\hat{\bd{b}}$ is the curvature vector of the field line, $\Omega_s=Z_s eB/m_s$ is the gyro-frequency, and $\avg{\varphi}_{\bd{R}}$ is the gyro-averaged electrostatic potential perturbation with fixed  $\bd{R}$. The electrostatic potential can be solved from the quasi-neutrality condition
\begin{equation}
    \sum_s\frac{Z_s^2 e^2 n_s}{T_s}\varphi=\sum_s Z_se\int \rmd^3\bd{v}\avg{h_s}_{\bd{r}},
\end{equation}
where $\avg{h_s}_{\bd{r}}$ is the gyro-averaged value of $h_s$ at fixed particle position  $\bd{r}=\bd{R}+\hat{\bd{b}}\times\bd{v}/\Omega_s$ and $\int \rmd^3 \bd{v}$ is the integration over velocity space.

Turbulent fluctuations are highly anisotropic, namely, they vary quickly across field lines but slowly along field lines. Therefore, one often uses the field-line following local coordinates $(x,y,\theta)$ defined as
\begin{equation}
    \nabla x=(rB_0)^{-1}\nabla\psi,\quad\nabla y=r\nabla\alpha, 
\end{equation}
and $\theta$ is the poloidal angle measured along the field lines at fixed $\alpha$. Here, we choose $B_0$ to be the on-axis magnetic-field strength and $r=\sqrt{2\psi/B_0}$.  The minor radius of the configuration is defined as $a=\sqrt{2\psi_{\rm edge}/B_0}$, where $\psi_{\rm edge}$ is the value of $\psi$ at the last flux surface. Note that existing local gyrokinetic simulations usually choose $a$ as the quantity \textsf{Aminor\_p} from the VMEC equilibria, and calculate $B_0=2\psi_{\rm edge}/a^2$ accordingly. Here, we choose a different convention in order to be more consistent with the NAE analysis in section~\ref{sec:theory_NAE} below. Also, the above definitions lead to $\nabla x=\nabla r$, so $x$ and $r$ are the same, but we will keep using $x$ in this section, consistent with the literature. In terms of the field-line following coordinates, the parallel derivative can be written as
\begin{equation}
    v_\parallel\hat{\bd{b}}\cdot\nabla=v_\parallel k_\theta\frac{\pd}{\pd\theta},
\end{equation}
where $k_\theta=(\pd l/\pd\theta)^{-1}$ and $l(\theta)$ is the distance measured along the field lines. The magnetic-drift term is
\begin{equation}
\label{eq:theory_magnetic_drift}
    \bd{v}_D\cdot\nabla h_s=\bd{v}_D\cdot\left[(\pd_x h_s)\nabla x+(\pd_y h_s)\nabla y\right].
\end{equation}
The nonlinear term is
\begin{equation}
\label{eq:theory_nonlinear}
    \frac{1}{B}\hat{\bd{b}}\times\nabla\avg{\varphi}_{\bd{R}}\cdot\nabla h_s=\frac{1}{B_0}\left(\frac{\pd\avg{\varphi}_{\bd{R}}}{\pd x}\frac{\pd h_s}{\pd y}-\frac{\pd\avg{\varphi}_{\bd{R}}}{\pd y}\frac{\pd h_s}{\pd x}\right),
\end{equation}
where the relation $\nabla x\times\nabla y=\nabla\psi\times\nabla\alpha/B_0=\bd{B}/B_0$ has been used. The equilibrium distribution function $F_s$ is often chosen as a local Maxwellian with density and temperature as flux functions, so that 
\begin{equation}
    \frac{\pd F_s}{\pd \psi}=-\frac{F_s}{B_0r}\left[\frac{1}{L_{ns}}+\left(\frac{\varepsilon}{T_s}-\frac{3}{2}\right)\frac{1}{L_{Ts}}\right],
\end{equation}
where $L_{ns}=(-\rmd\ln n_s/\rmd r)^{-1}$ and $L_{Ts}=(-\rmd\ln T_s/\rmd r)^{-1}$ are the length scales of the density and temperature gradients, respectively. 

\subsection{Geometric quantities from local gyrokinetic equations}
The local approach corresponds to a WKB approximation in $(x,y)$, so that fluctuating quantities can be written as
\begin{equation}
    A(x,y,\theta,\varepsilon,\mu,t)=\sum_{k_x,k_y}A_{k_x,k_y}(\theta,\varepsilon,\mu,t)\rme^{\rmi k_x x+\rmi k_y y}.
\end{equation}
The local gyrokinetic equation thus evolves the structure of each Fourier component in $(\theta,\varepsilon,\mu)$, and different Fourier components are coupled through the nonlinear term \eqref{eq:theory_nonlinear}. Under the WKB approximation, the gyro-average can be carried out in Fourier space; for example, for the potential:
\begin{equation}
    \avg{\varphi}_{\bd{R},k_x,k_y}=J_0(k_\perp\rho_s)\varphi_{k_x,k_y},
\end{equation}
where $J_0$ is the Bessel function, $\rho_s=\sqrt{2\mu B/m_s}/\Omega_s$ is the gyroradius, and 
\begin{equation}
\label{eq:theory_kperp}
    k_\perp=|k_x\nabla x+k_y\nabla y|=\sqrt{k_x^2|\nabla x|^2+2k_xk_y\nabla x\cdot\nabla y+k_y^2|\nabla y|^2}
\end{equation}
is the perpendicular wavenumber. Therefore, the local approach allows one to study turbulent transport in various magnetic geometries including tokamaks and stellarators, and geometric effects enter as 1-dimensional functions in $\theta$.  

For the study of transport levels from electrostatic ITG turbulence in this work, the local gyrokinetic equation involves 9 geometric quantities. Within local gyrokinetic codes such as GS2 \citep{Kotschenreuther95,Highcock12} and GX \citep{Mandell24}, these quantities are commonly referred to as \textsf{bmag}, \textsf{gradpar}, \textsf{cvdrift}, \textsf{gbdrift}, \textsf{cvdrift0}, 
\textsf{gbdrift0}, \textsf{gds2}, \textsf{gds21}, and \textsf{gds22}. The first two quantities are the magnetic field-strength normalized by $B_0$ and the inverse parallel connection length normalized by $a$, respectively:
\begin{equation}
    \textsf{bmag}=\frac{B}{B_0},\quad  \textsf{gradpar}=ak_\theta.
\end{equation} 
The next 4 quantities appear in the magnetic-drift term \eqref{eq:theory_magnetic_drift} and have also been made dimensionless using $B_0$ and $a$:
\begin{align}
    &\textsf{cvdrift}=B_0a\frac{\hat{\bd{b}}\times{\bd{\kappa}}\cdot\nabla y}{B},\quad \textsf{gbdrift}=B_0a\frac{\hat{\bd{b}}\times\nabla B\cdot\nabla y}{B^2},\\
    &\textsf{cvdrift0}=B_0a\frac{\hat{\bd{b}}\times{\bd{\kappa}}\cdot\nabla x}{B},\quad\textsf{gbdrift0}=B_0a\frac{\hat{\bd{b}}\times{\nabla B}\cdot\nabla x}{B^2},
\end{align}
The last 3 quantities arise from $k_\perp$ in the gyro-average operator:
\begin{equation}
    \textsf{gds2}=|\nabla y|^2,\quad \textsf{gds21}=\nabla x\cdot\nabla y,\quad \textsf{gds22}=|\nabla x|^2.
\end{equation}
Note that local gyrokinetic codes often include the magnetic shear 
\begin{equation}
    \hat{s}=-\frac{r}{\iota}\frac{\rmd\iota}{\rmd r}
\end{equation}
into the above definitions of \textsf{gbdrift0}, \textsf{cvdrift0}, \textsf{gds21}, and \textsf{gds22}, so that the ballooning angle $\theta_0=k_x/(k_y\hat{s})$ is used instead of $k_x$ \citep{Highcock12}. Since a coordinate transformation does not change the physical properties of the system, we do not include $\hat{s}$ here for simplicity. Also, it can be shown that \textsf{cvdrift0} and \textsf{gbdrift0} are identical due to the MHD force balance $\bd{J}\times\bd{B}=\nabla p$, where $p(\psi)$ is the equilibrium pressure and is assumed to be a flux function \citep{Highcock12}. Specifically, it can be written as
\begin{equation}
    \label{eq:theory_force_balance}\nabla_\perp\left(p+\frac{B^2}{2\mu_0}\right)=\frac{B^2}{\mu_0}\bd{\kappa},
\end{equation}
where $\nabla_\perp$ is the gradient perpendicular to $\bd{B}$. Crossing \eqref{eq:theory_force_balance} with $\nabla\psi$ and using $\nabla_\perp p=(\pd_\psi p)\nabla\psi$, one finds that \textsf{cvdrift0} and \textsf{gbdrift0} are the same. Therefore, within this study we will focus on 8 distinct geometric quantities. 

\subsection{Flux-tube equivalency for quasi-symmetric stellarators}
\label{sec:theory_flux_tube_equiv}
Stellarators are non-axisymmetric, meaning that the above geometric quantities generally depend on $\alpha$. For QS stellarators, however, the magnetic-field strength can be written as $B=B(\psi,\vartheta)$ in Boozer coordinates with $\vartheta=\theta-N\zeta$, so that
\begin{equation}
    \textsf{bmag} =\frac{B(\psi,\vartheta)}{B_0}
\end{equation}
is independent of $\alpha$ in the field-line following coordinates $(\psi,\vartheta,\alpha)$. Moreover,  4 other quantities \textsf{gradpar}, \textsf{cvdrift}, \textsf{gbdrift}, and \textsf{cvdrift0} are $\alpha$-independent too \citep{NiesProposal}. To show this, we write the gradient of a scalar $f$ in field-line following coordinates as $\nabla f(\psi,\alpha,\vartheta)=(\pd_\psi f)\nabla\psi+(\pd_\alpha f)\nabla\alpha+(\pd_\vartheta f)\nabla\vartheta$, so that
\begin{equation}
    \bd{B}\cdot\nabla f=(\nabla\psi\times\nabla\alpha\cdot\nabla\vartheta)\frac{\pd f}{\pd\vartheta}.
\end{equation}
Since $\alpha=\theta-\iota\zeta=\vartheta-\iota_N\zeta$ with $\iota_N=\iota-N$, we have
\begin{equation}
    \textsf{gradpar}=\frac{\iota_N a}{B\sqrt{g}},
\end{equation}
where
\begin{equation} \sqrt{g}=\left(\nabla\psi\times\nabla\vartheta\cdot\nabla\zeta\right)^{-1}=\frac{G(\psi)+\iota I(\psi)}{B^2}
\end{equation}
is the Jacobian in Boozer coordinates. Therefore, within the flux surface, \textsf{gradpar}  depends only on $B$ and hence is $\alpha$-independent. The same can be shown for \textsf{gbdrift0} and $\textsf{gbdrift}$: write the magnetic field as
\begin{equation}
    \bd{B}=G_N(\psi)\nabla\zeta+I(\psi)\nabla\vartheta+\delta(\psi,\vartheta)\nabla\psi,\quad G_N=G+NI,
\end{equation}
where $\delta$ no longer depends on $\zeta$ due to QS. Since $\nabla B=(\pd_\psi B)\nabla\psi+(\pd_\vartheta B)\nabla\vartheta$, we have
\begin{equation}
\bd{B}\times\nabla B\cdot\nabla\psi=(G_N\nabla\zeta+I\nabla\vartheta+\delta\nabla\psi)\cdot\pd_\vartheta B \nabla\vartheta\times\nabla\psi=-\frac{G_N}{\sqrt{g}}\pd_\vartheta B,
\end{equation}
so that
\begin{equation}
\textsf{gbdrift0}=-\frac{a}{rB^3}\frac{G_N}{\sqrt{g}}\frac{\pd B}{\pd\vartheta}
\end{equation}
is $\alpha$-independent. We also have
\begin{align}
    \bd{B}\times\nabla B\cdot\nabla\alpha&=(G_N\nabla\zeta+I\nabla\vartheta+\delta\nabla\psi)\cdot(\pd_\psi B\nabla\psi+\pd_\vartheta B\nabla\vartheta)\times\nabla(\vartheta-\iota_N\zeta)
    \notag
    \\
    &=\frac{1}{\sqrt{g}}\left[\left(G_N+\iota_N I\right)\frac{\pd B}{\pd\psi}+\left(\frac{\rmd \iota_N}{\rmd \psi}\zeta G_N-\iota_N \delta\right)\frac{\pd B}{\pd\vartheta}\right],
\end{align}
so that
\begin{equation}
    \textsf{gbdrift}=\frac{B_0ar}{B^3}\frac{1}{\sqrt{g}}\left[\left(G_N+\iota_N I\right)\frac{\pd B}{\pd\psi}+\left(\frac{\rmd \iota_N}{\rmd \psi}\zeta G_N-\iota_N \delta\right)\frac{\pd B}{\pd\vartheta}\right].
\end{equation}
Note that here \textsf{gbdrift} depends on $\alpha$ through $\zeta=(\vartheta-\alpha)/\iota_N$, but since the magnetic-drift term \eqref{eq:theory_magnetic_drift} involves only the linear combination $k_x\cdot \textsf{gbdrift0}+k_y\cdot\textsf{gbdrift}$, the $\alpha$-dependence in \textsf{gbdrift} can be removed by a coordinate transform $k_x\to k_x-\alpha k_y\hat{s}\iota/\iota_N$. The Fourier triad condition and the coupling coefficients in the nonlinear term \eqref{eq:theory_nonlinear} remain unchanged under this coordinate transform. Finally, from the MHD force balance \eqref{eq:theory_force_balance}, we have
\begin{equation}
    \textsf{cvdrift}=\textsf{gbdrift}+\frac{B_0ar}{B^2}\mu_0\frac{\rmd p(\psi)}{\rmd \psi}.
\end{equation}
The difference comes from the scalar pressure gradient and is $\alpha$-independent. Therefore, under the same coordinate transform $k_x\to k_x-\alpha k_y\hat{s}\iota/\iota_N$, \textsf{cvdrift} is also $\alpha$-independent. 

It is well known that the neoclassical-transport properties of QS stellarators are isomorphic to those of tokamaks \citep{Boozer83}. For the calculation of neoclassical transport, only three quantities are involved: \textsf{bmag}, \textsf{gradpar}, and \textsf{cvdrift0}, which do not depend on field lines. For turbulent transport, two additional quantities \textsf{gbdrift} and \textsf{cvdrift} are also $\alpha$-independent. (To our knowledge such ``flux-tube equivalency'' has not been explicitly mentioned until \cite{NiesProposal}.) Note however that for turbulent transport, the 3 other quantities \textsf{gds2}, \textsf{gds21}, and \textsf{gds22} are still $\alpha$-dependent, so that different field lines should still exhibit different transport levels. The additional effects from the $\alpha$-dependent quantities will be discussed in section~\ref{sec:numerical}.
\subsection{Describing stellarators and tokamaks from the near-axis expansion}
\label{sec:theory_NAE}
The NAE framework provides a systematic approach to constructing QS stellarator configurations \citep{Garren91a,Garren91b,Landreman19a,Landreman19b,Jorge20,Rodriguez22,Rodriguez23}.  Given a magnetic axis $\bd{r}_0(\zeta)$, we can calculate  its arc length $l(\zeta)=\int|\rmd\bd{r}_0/\rmd\zeta|\rmd\zeta$, curvature $\kappa(\zeta)$, and torsion $\tau(\zeta)$. We can also define orthonormal vectors along the axis, which are the tangent vector $\hat{\bd{t}}(\zeta)$, the normal vector $\hat{\bd{n}}(\zeta)$, and the binormal vector $\hat{\bd{\rmb}}(\zeta)$. These quantities are obtained through the following relations \citep{Mercier64,Landreman19b}:
\begin{equation}
\hat{\bd{t}}=\frac{\rmd \bd{r}_0}{\rmd l},\quad\kappa\hat{\bd{n}}=\frac{\rm d\hat{\bd{t}}}{\rmd l},\quad \hat{\bd{\rmb}}=\hat{\bd{t}}\times\hat{\bd{n}},\quad \tau\hat{\bd{n}}=-\frac{\rmd\hat{\bd{\rmb}}}{\rmd l},
\end{equation}
where $\rmd/\rmd l=(\rmd l/\rmd \zeta)^{-1}\rmd/\rmd\zeta$. Specifically, we obtain $\hat{\bd{t}}$ from the first equation (which by definition satisfies $|\hat{\bd{t}}|=1$), $\kappa$ and $\hat{\bd{n}}$ from the second equation assuming $\kappa>0$ and $|\hat{\bd{n}}|=1$, $\hat{\bd{\rmb}}$ from the third equation, and $\tau$ from the last equation. This procedure can be carried out when $\kappa$ does not vanish anywhere, which applies to the QA and QH configurations \citep{Landreman18}. Then, the position vector $\bd{r}$ can be expressed in Boozer coordinates as
\begin{equation}
    \bd{r}(r,\vartheta,\zeta)=\bd{r}_0(\zeta)+X(r,\vartheta,\zeta)\hat{\bd{n}}(\zeta)+Y(r,\vartheta,\zeta)\hat{\bd{\rmb}}(\zeta)+Z(r,\vartheta,\zeta)\hat{\bd{t}}(\zeta).
\end{equation}
Here, the helical angle $\vartheta=\theta-N\zeta$ is used so that the following descriptions apply to both QA ($N=0$) and QH ($N\neq 0$) configurations.

The NAE corresponds to expanding the above quantities in $r$. For our calculations, the relevant quantities are listed as follows. The magnetic field is written as
\begin{equation}
    B(r,\vartheta,\zeta)=B_0(\zeta)+rB_1(\vartheta,\zeta)+r^2B_2(\vartheta,\zeta)+\mc{O}(r^3),
\end{equation}
where the coefficients can be further expanded as
\begin{equation}
    B_1=B_{1\rms}(\zeta)\sin\vartheta+B_{1\rmc}(\zeta)\cos\vartheta,\quad B_2=B_{20}(\zeta)+B_{2\rms}(\zeta)\sin 2\vartheta+B_{2\rmc}(\zeta)\cos 2\vartheta.
\end{equation}
The function $\delta$ in \eqref{eq:theory_covariant_B}, which is the radial covariant component of $\bd{B}$, can also be expanded as
\begin{equation}
    \delta(r,\vartheta,\zeta)=\delta_0(\zeta)+r\delta_1(\vartheta,\zeta)+\mc{O}(r^2),\quad \delta_1=\delta_{1\rms}(\zeta)\sin\vartheta+\delta_{1\rmc}(\zeta)\cos\vartheta.
\end{equation}
The flux functions $G$, $I$, $p$, and $\iota$ are assumed to be analytic in $\psi=B_0 r^2/2$, so that they can be expanded in even powers of $r$,
\begin{align}
    G(r)=G_0+r^2G_2+\mc{O}(r^4),\quad I(r)=r^2I_2+\mc{O}(r^4),
    \\
     p(r)=p_0+r^2p_2+\mc{O}(r^4),\quad \iota(r)=\iota_0+r^2\iota_2+\mc{O}(r^4),
\end{align}
where $I_0=0$ assuming there is no current singularity on the axis. In the next section, the geometric quantities outlined in section~\ref{sec:theory_flux_tube_equiv} will be calculated order by order, based on which the  equivalent-tokamak theory will be developed.
\subsection{Describing equivalent tokamaks as local Miller equilibria}
\label{sec:theory_miller}
The Miller equilibrium is a widely used local analytic equilibrium model for tokamak plasmas \citep{Miller98}. The model describes flux surfaces with the following form in cylindrical coordinates:
\begin{equation}
\label{eq:theory_miller}
    R=R_0(\rho)+\rho\cos\lbrace\theta+[\sin^{-1}\Delta(\rho)]\sin\theta\rbrace,\quad z=\varkappa(\rho)\rho\sin\theta.
\end{equation}
Here, $\rho$ and $\theta$ are the minor radius and the poloidal angle, and $\varkappa$ and $\Delta$ describe the elongation and triangularity of the flux surface, respectively. Given $R_0$, $\varkappa$, $\Delta$, and their radial derivatives $R_0'$, $\varkappa'$, and $\Delta'$ (we use the prime to denote the derivative with respect to $\rho$), one can expand the Grad--Shafranov equation around the flux surface and obtain a local MHD equilibrium with the following additional parameters: $q=\iota^{-1}$, $\hat{s}$, and $\beta'$ with $\beta=2\mu_0 p/B_0^2$. After the local Grad-Shafranov equation is solved, the geometric quantities listed in section~\ref{sec:theory_flux_tube_equiv} can be calculated following the steps presented, for example, in the appendices of \cite{Gaur23}. 

The relevant parameters for the Miller model are listed in table~\ref{table:miller}. We note that a local Miller equilibrium does not determine a corresponding global MHD equilibrium. However, given a global MHD equilibrium, one can find a set of Miller parameters that provide a good fit to a flux surface, which can be done numerically using codes such as \textsf{pyrokinetics} \citep{pyrokinetics}. Also, the Miller model can only describe up-down symmetric flux surfaces, so the geometric coefficients are either odd or even in $\theta$. However, for QS stellarators with stellarator symmetry \citep{Dewar98}, their geometric quantities will satisfy the same properties, so the Miller model will suffice.

\begin{table}
  \begin{center}
\def~{\hphantom{0}}
  \begin{tabular}{lccccccccccc}
      parameters & $\rho$ & $R_0$ & $R_0'$  & $\varkappa$ & $\varkappa'$ & $\Delta$ & $\Delta'$ & $q$& $\hat{s}$&$\beta'$
      \\
      \textsf{GX} variables& \textsf{rhoc} & \textsf{Rmaj}&  \textsf{shift} & \textsf{akappa} & \textsf{akappri} & \textsf{tri} & \textsf{tripri}& \textsf{qinp} & \textsf{shat} & \textsf{betaprim}
      \\
      $\beta=2.5\%$ QA& 0.5&10.2&0.20&1.56&-0.40&-0.10&-0.17&0.68&-0.02& -0.055
      \\
      $\beta=2.5\%$ QH& 0.5&5.00&0.15&0.98&-0.25&-0.16&-0.21&1.55&-0.45& -0.053
  \end{tabular}
  \caption{The Miller parameters and their optimized values for the equivalent tokamaks discussed in section~\ref{sec:application}. Since we use the geometry module from \textsf{GX} for the optimization, the corresponding variable names in the code are also listed for reference.}
  \label{table:miller}
  \end{center}
\end{table}
\section{Derivation of equivalent tokamaks of quasi-symmetric stellarators from the near-axis expansion}
\label{sec:theory}
\subsection{Calculating geometric quantities from the near-axis expansion}
For QS stellarators, the NAE coefficients can be greatly simplified \citep{Landreman19b}. Assuming stellarator symmetry \citep{Dewar98}, the coefficients $B_{1\rms}$, $B_{2\rms}$, and $\delta_{1\rmc}$ are zero. Then, when solving the NAE system up to second order in $r$, the input parameters of the NAE model are the axis shape $\bd{r}_0$, the on-axis field strength $B_0$, the coefficients $I_2$, $p_2$, and $B_{2\rmc}$, and another free parameter $\bar{\eta}$. The toroidal angle $\zeta$ is defined such that $R_0=\rmd l/\rmd\zeta$ is a constant; i.e., $\zeta$ measures the arc length along the axis and $|R_0|$ is the averaged major radius of the axis (note that $R_0$ can be either positive or negative depending on the direction of $\zeta$). Then, 
\begin{equation}
    G_0=B_0R_0.
\end{equation} 
The on-axis rotational transform $\iota_0$ is determined as the eigenvalue from the following Riccati equation:
\begin{equation}
\label{eq:NAE_sigma}
    \frac{\rmd\sigma}{\rmd\zeta}+\iota_{N0}\left(1+\sigma^2+\frac{\etabar^4}{\kappa^4}\right)-\frac{2G_0\etabar^2}{B_0\kappa^2}\left(\frac{I_2}{B_0}-\tau\right)=0,\quad\iota_{N0}=\iota_0-N,
\end{equation}
and the solution satisfies  $\sigma(0)=\sigma(2\pi)=0$ assuming stellarator symmetry. For tokamaks with circular magnetic axes, $|R_0|$ is the major radius, $\kappa=R_0^{-1}$, and $N=\tau=0$. The parameter $\bar{\eta}$ can still be freely chosen for tokamaks, and flux surfaces are circular with $\bar{\eta}=\kappa$ and elliptical otherwise. Then, the Riccati equation~\eqref{eq:NAE_sigma}  simplifies to
\begin{equation}
 \label{eq:NAE_sigma_tokamak}    \iota_0\left(1+\frac{\etabar^4}{\kappa^4}\right)=\frac{2G_0\etabar^2}{B_0\kappa^2}\frac{I_2}{B_0},
\end{equation}
so that $\iota_0$ can be obtained algebraically.

Including terms up to second order, the magnetic-field strength can be written as
\begin{equation}
     B=B_0(1+\bar{\eta}r\cos\vartheta)+r^2(B_{20}+B_{2\rmc}\cos2\vartheta),
\end{equation}
namely, 
\begin{equation}
   B_{1\rmc}=B_0\bar{\eta}.  
\end{equation}
While $B_{2\rmc}$ is a free $\zeta$-independent parameter in the model, $B_{20}$ is solved from an ordinary differential equation in $\zeta$ so that it is not guaranteed to be $\zeta$-independent; good QS properties can be achieved only when the dependence of $B_{20}$ on $\zeta$ becomes small enough \citep{Rodriguez23,Landreman19b}. For tokamaks with circular axes, the differential equation becomes algebraic so that $B_{20}$ is indeed $\zeta$-independent and can be solved analytically. Following the outline from \cite{Landreman19b}, $B_{20}$ is obtained as
\begin{align}
\notag
B_{20}^\rmt(\bar{\eta},I_2,p_2,B_{2\rmc})
=\;&
-3\,\frac{\bar{\eta}^{4}-1}{\bar{\eta}^{4}-3}\,B_{2\rmc} \\
\notag
&+
\frac{\bar{\eta}^{2}\left(2\bar{\eta}^{12}-3\bar{\eta}^{8}-12\bar{\eta}^{4}-7
+8I_{2}^{2}\bar{\eta}^{4}(2-\bar{\eta}^{4})\right)}
{2(\bar{\eta}^{4}-3)(\bar{\eta}^{4}+1)^{2}} \\
&+
\mu_{0}p_{2}\,
\frac{(\bar{\eta}^{4}+1)^{2}-2I_{2}^{2}(\bar{\eta}^{4}-3)}
{2I_{2}^{2}(\bar{\eta}^{4}-3)}.\label{eq:NAE_B20}
\end{align}
Here, the superscript ``t'' refers to tokamaks, and we have assumed $R_0=B_0=1$ for simplicity. (The details are presented in Appendix~\ref{app:B20}.)  The coefficients $G_2$ and $\delta_{1\rms}$ can be obtained from MHD force balance as
\begin{equation}
\label{eq:NAE_MHD}
    G_2=-\iota_0I_2-\frac{\mu_0p_2 G_0}{B_0^2},\quad\delta_{1\rms}=-\frac{4\mu_0 p_2G_0\bar{\eta}}{\iota_{N0} B_0^3}.
\end{equation}
Finally, the coefficient $\iota_2$ is also nonzero and is related to the magnetic shear as
\begin{equation}
    \hat{s}=-\frac{r}{\iota}\frac{\rmd\iota}{\rmd r}\approx-\frac{2\iota_2 r^2}{\iota_0}.
\end{equation}
However, $\iota_2$ cannot be obtained unless we solve the third-order equations of the NAE. Since the third-order equations are complicated and usually overdetermined, we do not seek to determine $\iota_2$ from the NAE framework. Consequently, in the following $\hat{s}$ is treated as a  given $\mc{O}(r^0)$ parameter rather than a $\mc{O}(r^2)$ quantity from the NAE expansion.

With the above NAE coefficients determined, the geometric quantities are thus calculated as follows, keeping terms up to $\mc{O}(r)$:
\begin{align}
\label{eq:NAE_bmag_gradpar}
&\mathsf{bmag}=1+\bar{\eta}r\cos\vartheta,\quad
\mathsf{gradpar}=\frac{a\iota_{N0}(1+\bar{\eta}r\cos\vartheta)}{R_0},
\\
\label{eq:NAE_gbdrift0}
&\mathsf{gbdrift0}=\mathsf{cvdrift0}=\frac{a\bar{\eta}\sin\vartheta}{1+\bar{\eta}r\cos\vartheta}+\frac{2ar B_{2\rmc}\sin2\vartheta}{B_0(1+\bar{\eta}r\cos\vartheta)},
\\
\label{eq:NAE_gbdrift}
&\mathsf{gbdrift}=\frac{a\bar{\eta}(\cos\vartheta+\hat{s}_N \vartheta\sin\vartheta)}{1+\bar{\eta}r\cos\vartheta}
+\frac{2ar\left[B_{20}+B_{2\rmc}(\cos2\vartheta+\hat{s}_N\vartheta\sin 2\vartheta)\right]}{B_0(1+\bar{\eta}r\cos\vartheta)},
\\
\label{eq:NAE_cvdrift}
&\mathsf{cvdrift}=\mathsf{gbdrift}+\frac{2\mu_0p_2 ar}{B_0^2(1+\bar{\eta}r\cos\vartheta)^2},
\end{align}
with
\begin{gather}
    \hat{s}_N=-\frac{r}{\iota_N}\frac{\rmd\iota_N}{\rmd r}\approx\frac{\iota_0}{\iota_{N0}}\hat{s}.
\end{gather}
The other coefficients $G_2$, $I_2$, and $\delta_{1\rms}$ appear only in the $\mc{O}(r^2)$ terms and are not included.
\subsection{Equivalent tokamaks with zeroth-order accuracy}
\label{subsec:theory_zeroth}
Given a QS stellarator configuration, we wish to find a tokamak whose geometric quantities are the same. From \eqref{eq:NAE_bmag_gradpar} to \eqref{eq:NAE_cvdrift}, keeping only $\mc{O}(r^0)$ terms, the equivalent tokamak is found to satisfy the following relation:
\begin{equation}
\label{eq:NAE_zeroth_order}
     a^\rmt=\frac{\bar{\eta}^\rms}{\bar{\eta}^\rmt}a^\rms,
     \quad
     \iota_0^\rmt=\frac{R_0^\rmt}{R_0^\rms}\frac{\bar{\eta}^\rmt}{\bar{\eta}^\rms}\iota_{N0}^\rms,
     \quad
     \hat{s}^\rmt=\hat{s}_N^\rms.
\end{equation}
Here, the superscripts ``t'' and ``s''  refer to tokamaks and stellarators, and the subscript ``$N$'' has been omitted for tokamaks. For the tokamak, both the major radius $R_0^\rmt$ and the parameter $\bar{\eta}^\rmt$ can be freely chosen, and the flux-surface shape is circular if  $R_0^\rmt=1/\bar{\eta}^\rmt$ (and elliptical otherwise).  Then, the minor radius $a^\rmt$ and the on-axis rotational transform $\iota_0^\rmt$ are calculated according to the above equation. To achieve such $\iota_0^\rmt$, the parameter $I_2^\rmt$ must be chosen according to \eqref{eq:NAE_sigma_tokamak}, which leads to
\begin{equation}
    I_2^\rmt=\frac{B_0^\rmt\iota_0^\rmt\left[1+(R_0^{\rmt}\bar{\eta}^\rmt)^4\right]}{2(R_0^\rmt)^3(\etabar^\rmt)^2}.
\end{equation}
The magnetic shear $\hat{s}^\rmt$ is not a free parameter in the NAE and cannot be easily calculated, but it does not enter the first- or the second-order NAE equations. In other words, we can consider $\hat{s}^\rmt$ as a prescribed quantity.

A simple choice of the equivalent tokamak with zeroth-order accuracy is $\bar{\eta}^\rmt=\bar{\eta}^\rms$ and $R_0^\rmt=1/\bar{\eta}^\rmt=1/\bar{\eta}^\rms$. Then, the equivalent tokamaks have circular flux surfaces with the same minor radius $a^\rmt=a^\rms$, and are characterized by three parameters: the aspect ratio, the rotational transform, and the magnetic shear. Their respective values are given as
\begin{equation}
\label{eq:NAE_zeroth_order_circ}
    \frac{
    R_0^\rmt
    }{a^\rmt}=\frac{1}{a^\rms\bar{\eta}^\rms},\quad\iota_0^\rmt=\frac{\iota_{N0}^\rms}{\bar{\eta}^\rms R_0^\rms}\quad
    \hat{s}^\rmt=\hat{s}_N^\rms.
\end{equation}
Therefore, when compared with the QS stellarator, the equivalent tokamak has a different aspect ratio, a different rotational transform, and a different magnetic shear (if $N\neq 0$).

\subsection{Equivalent tokamaks with first-order accuracy}
With $a^\rmt$, $\iota_0^\rmt$, and $\hat{s}^\rmt$ determined from \eqref{eq:NAE_zeroth_order}, we wish to choose other parameters so that the $\mc{O}(r^1)$ terms from \eqref{eq:NAE_bmag_gradpar} to \eqref{eq:NAE_cvdrift} also match between the tokamak and the QS stellarator. For simplicity, we assume $B_0^\rmt=B_0^\rms$ and $R_0^\rmt=R_0^\rms$. This leads to
\begin{equation}
    \bar{\eta}^\rmt r^\rmt=\bar{\eta}^\rms r^\rms,\quad p_2^\rmt a^\rmt r^\rmt=p_2^\rms a^\rms r^\rms,\quad B_{2\rmc}^\rmt a^\rmt r^\rmt=B_{2\rmc}^\rms  a^\rms r^\rms,\quad B_{20}^\rmt a^\rmt  r^\rmt=B_{20}^\rms a^\rms r^\rms.
\end{equation}
Using the results from \eqref{eq:NAE_zeroth_order}, we have
\begin{equation}
    r^\rmt=\left(\frac{{\bar{\eta}^\rms}}{{\bar{\eta}^\rmt}}\right)r^\rms,\quad p_2^\rmt=\left(\frac{\bar{\eta}^\rmt}{\bar{\eta}^\rms}\right)^2p_2^\rms,\quad B_{2\rmc}^\rmt=\left(\frac{\bar{\eta}^\rmt}{\bar{\eta}^\rms}\right)^2B_{2\rmc}^\rms,\quad B_{20}^\rmt=\left(\frac{\bar{\eta}^\rmt}{\bar{\eta}^\rms}\right)^2B_{20}^\rms.
\end{equation}
Therefore, at first-order accuracy, we look at the flux surface with the same normalized minor radius $r^\rmt/a^\rmt=r^\rms/a^\rms$ as the QS stellarator, and the parameters $p_2^\rmt$ and $B_{2\rmc}^t$ are chosen according to the above equations. However, $B_{20}$ is not a free parameter and is determined from \eqref{eq:NAE_B20}. To match $B_{20}$, we require
\begin{equation}
    \label{eq:NAE_B20_match}
    B_{20}^\rmt(\bar{\eta}^\rmt,I_2^\rmt,p_2^\rmt,B_{2\rmc}^\rmt)=\left(\frac{\bar{\eta}^\rmt}{\bar{\eta}^\rms}\right)^2B_{20}^\rms.
\end{equation}
Equation \eqref{eq:NAE_B20_match} does not always have a solution for $\bar{\eta}^\rmt$, but as will be discussed in section \ref{sec:application}, for some QS stellarators we can find $\bar{\eta}^\rmt$ that meet the requirement \eqref{eq:NAE_B20_match}. Since the solutions of $\bar{\eta}^\rmt$ do not always equal $1/R^\rmt$, the resulting equivalent tokamaks no longer have circular flux surfaces and their $a^\rmt$ and $\iota_0^\rmt$ are given by \eqref{eq:NAE_zeroth_order} instead of \eqref{eq:NAE_zeroth_order_circ}. We also note that $\hat{s}^\rmt$ can still not be determined at this order, and therefore is not guaranteed to match between the QS stellarator and the equivalent tokamak. 
\section{Application to several  quasi-symmetric configurations}
\label{sec:application}
\subsection{Vacuum configurations}
\label{sec:application_precise}
\begin{figure}
    \centering
    \includegraphics[width=0.8\linewidth]{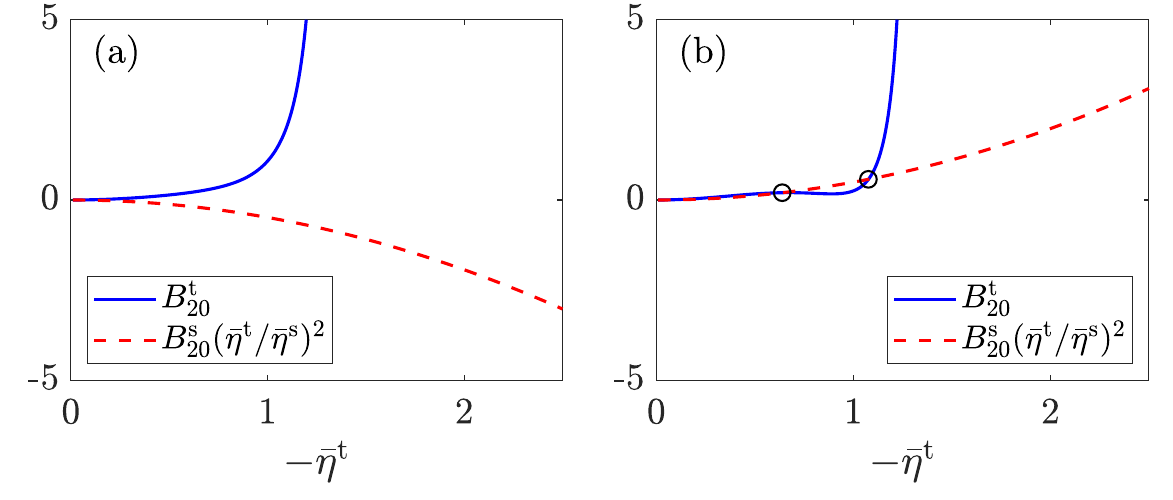}
    \caption{The match of $B_{20}$ (relation \eqref{eq:NAE_B20_match}) for the vacuum QS configurations from \cite{Landreman22}, where all quantities have been normalized to $R_0=B_0=1$. (a): The precise QA configuration with $a=0.153$, $\bar{\eta}=-0.72$, $\iota=0.42$, $\hat{s}=0.01$, $B_{2\rmc}=0.3$, $B_{20}=-0.25$, $p_2=0$, and $N=0$. No solution can be found. (b) The precise QH configuration with $a=0.087$, $\bar{\eta}=-1.95$, $\iota=-1.24$, $\hat{s}=-0.0014$, $B_{2\rmc}=0.66$, $B_{20}=1.89$, $p_2=0$, and $N=-4$. The two solutions $\bar{\eta}=-0.64$ and  $\bar{\eta}^\rmt=-1.08$ are highlighted by the black circles.}
    \label{fig:precise_QS_B20}
\end{figure}
As the first application, we find equivalent tokamaks for the precise QA and the precise QH configurations from \cite{Landreman22}. They are vacuum configurations and can be accurately described from the NAE, and their relevant parameters are described in the caption of figure~\ref{fig:precise_QS_B20}. To the zeroth order, the equivalent tokamaks are circular and are characterized by 3 quantities:
\begin{align}
\label{eq:application_precise_QA}
    &{\rm QA}: \bigg|\frac{R_0^\rmt}{a^\rmt}\bigg|=9.08,\quad
    \iota^\rmt=-0.59,\quad \hat{s}^\rmt=0.01, 
    \\
\label{eq:application_precise_QH}
    &{\rm QH}: \bigg|\frac{R_0^\rmt}{a^\rmt}\bigg|=5.91,\quad \iota^\rmt=-1.41, \quad
    \hat{s}^\rmt= 0.0006.    
\end{align}
Note that the QH has a much larger $|\bar{\eta}^\rms|$ compared to the QA, so its equivalent tokamak has a smaller $|R_0^\rmt/a^\rmt|$. In other words, the equivalent tokamak for the QH is more compact, in contrast to the original configurations ($R_0^\rms/a^\rms=6.5$ for the QA and $R_0^\rms/a^\rms=11.5$ for the QH). The equivalent tokamak for the QH also has a much larger $|\iota^\rmt|$ compared to the QA, which mainly comes from the transform $\iota_N=\iota-N$ with $N=-4$. Finally, both cases have very small $\hat{s}^\rmt$, a common feature of vacuum stellarator configurations.

\begin{figure}
    \centering
    \includegraphics[width=0.7\linewidth]{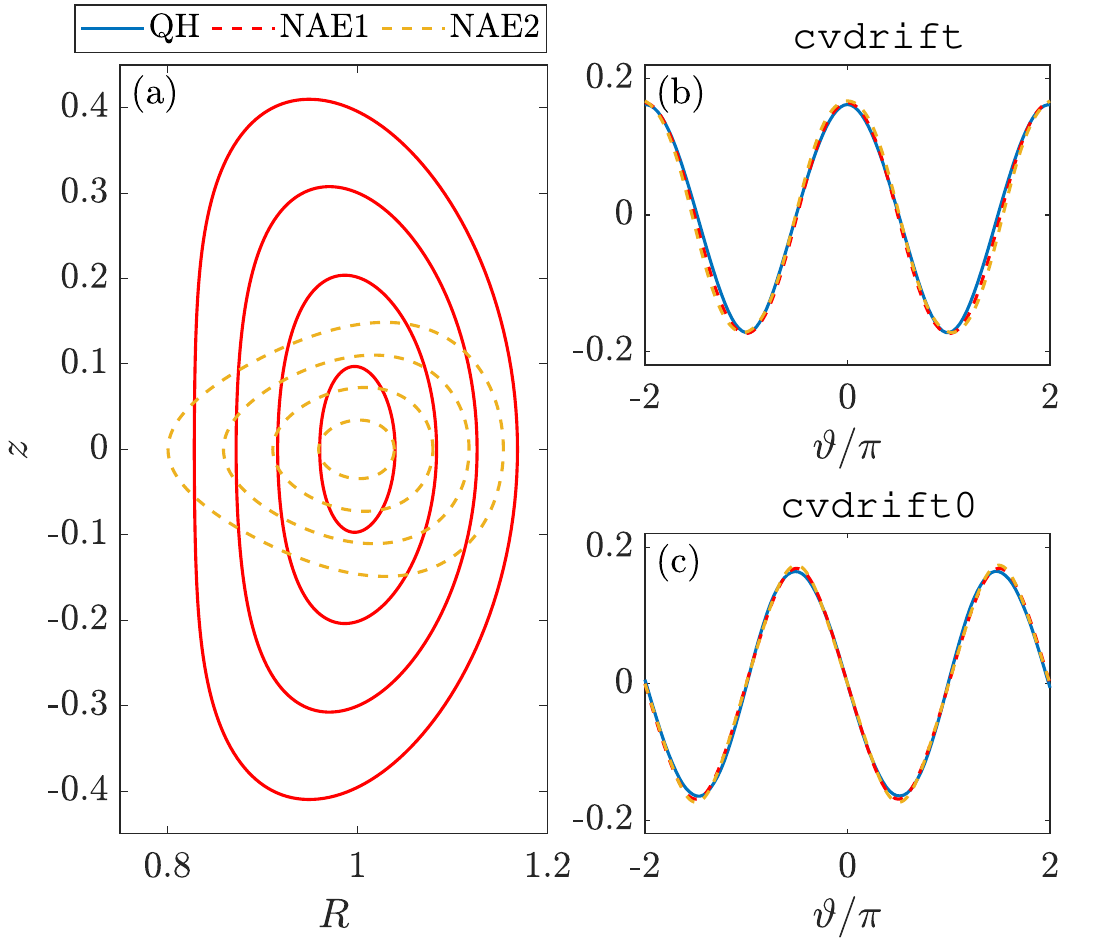}
    \caption{Equivalent tokamaks for the precise QH configuration from \cite{Landreman22}. (a) The flux surfaces of the global MHD equilibria from \textsf{DESC}, which are consistent with the NAE description near the axis. Here, ``NAE1'' and ``NAE2'' refer to the $\etabar=-0.64$ and the $\etabar=-1.08$ solutions, respectively. (b) Comparison of \textsf{cvdrift} between the stellarator (blue solid lines) and \textsf{DESC} results (red dashed lines and yellow dashed-dotted lines) at $r/a=0.5$. (c) Comparison of \textsf{cvdrift0} at $r/a=0.5$.}
    \label{fig:equiv_precise_QH}
\end{figure}

To find equivalent tokamaks with first-order accuracy, the relation \eqref{eq:NAE_B20_match} must be satisfied, and the results are shown in figure~\ref{fig:precise_QS_B20}. For the precise QA configuration, no solution of $\etabar^\rmt$ can be found. For the precise QH configuration, two solutions are found at $\bar{\eta}=-0.64$ and $\bar{\eta}^\rmt=-1.08$, and  other parameters are (with $B_0^t=R_0^t=1$ and $p_2^t=0$)
\begin{align}
\label{eq:application_precise_QH1}
    &\etabar=-0.64: \bigg|\frac{R_0^\rmt}{a^\rmt}\bigg|=3.77,~ \iota^\rmt=-0.90,~ I_2^\rmt=-1.29,~ B_{2\rmc}^\rmt=0.07,~ B_{20}^\rmt=0.20,
    \\
\label{eq:application_precise_QH2}
    &\etabar=-1.08: \bigg|\frac{R_0^\rmt}{a^\rmt}\bigg|=6.37,~ \iota^\rmt=-1.52,~ I_2^\rmt=-1.54,~ B_{2\rmc}^\rmt=0.20,~ B_{20}^\rmt=0.57.
\end{align}
Here, we have assumed $R_0^\rmt=B_0^\rmt=1$ for the equivalent tokamak. To generate the global MHD equilibrium for the equivalent tokamak, we use the code \textsf{DESC}, which has been recently upgraded with the capability to generate equilibria that are consistent with the NAE description near the axis \citep{pyqsc,Panici26}. The flux surfaces of the equivalent tokamaks are shown in figure~\ref{fig:equiv_precise_QH}(a), and we plot \textsf{cvdrift} and \textsf{cvdrift0} at $r/a=0.5$ in figures~\ref{fig:equiv_precise_QH}(b,c), which show good agreement. It is also straightforward to verify from \eqref{eq:application_precise_QH1} and \eqref{eq:application_precise_QH2} that \textsf{bmag} and \textsf{gradpar} should match between the stellarator and their equivalent tokamaks.

\subsection{Configurations with finite pressure}
\label{subsec:application_LBD}
\begin{figure}
    \centering
    \includegraphics[width=0.8\linewidth]{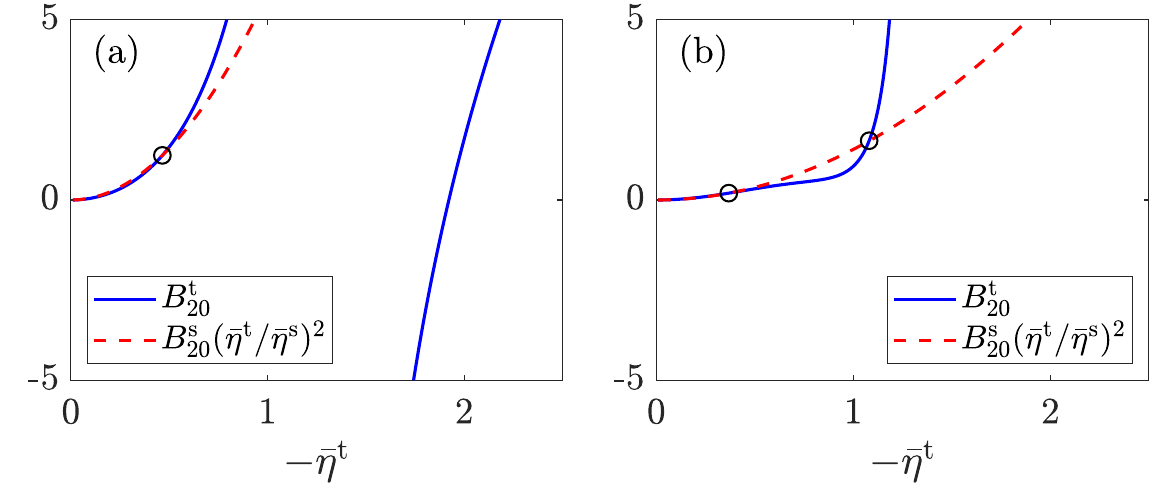}
    \caption{The match of $B_{20}$ (relation \eqref{eq:NAE_B20_match}) for the $\beta=2.5\%$ QS configurations from \cite{LBD22}, where all quantities have been normalized to $R_0=B_0=1$. (a): The $\beta=2.5\%$ QA configuration with $a=0.159$, $\bar{\eta}=-0.67$, $\iota=0.40$, $\hat{s}=-0.45$, $B_{2\rmc}=-0.39$, $B_{20}=2.57$, $p_2=-9.47\times 10^5$, and $N=0$. The solution is $\bar{\eta}^\rmt=-0.47$ and is highlighted by the black circle. (b) The $\beta=2.5\%$ QH configuration with $a=0.11$, $\bar{\eta}=-1.80$, $\iota=-1.09$, $\hat{s}=0.074$, $B_{2\rmc}=0.89$, $B_{20}=4.56$, $p_2=-1.85\times 10^6$, and $N=-4$. The two solutions $\bar{\eta}=-0.37$ and  $\bar{\eta}^\rmt=-1.08$ are highlighted by the black circles.}
    \label{fig:LBD_QS_B20}
\end{figure}

\begin{figure}
    \centering
    \includegraphics[width=1\linewidth]{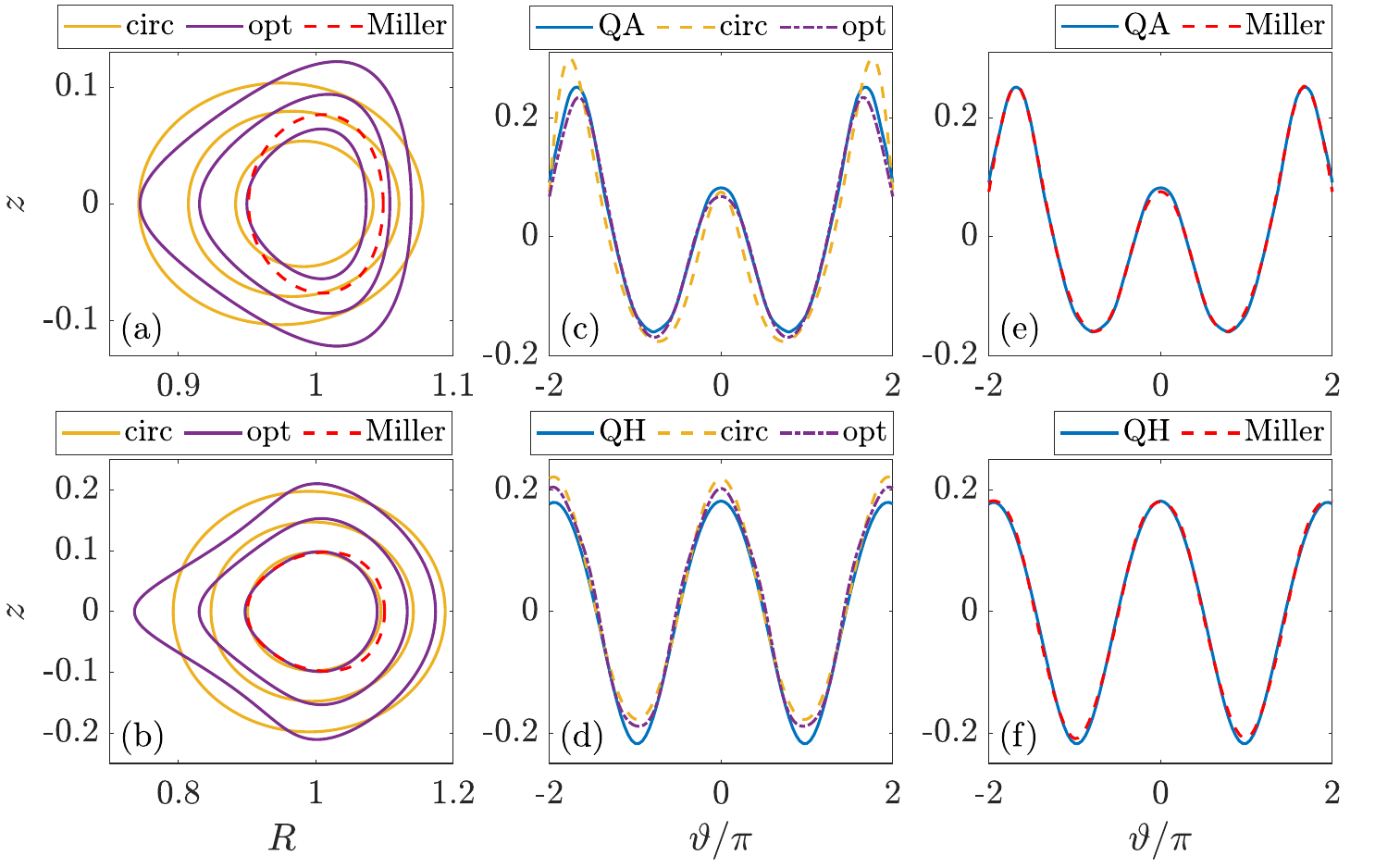}
    \caption{Equivalent tokamaks for the $\beta=2.5\%$ QA and QH configurations from \cite{LBD22}. (a,b) The flux surfaces of the global MHD equilibria from \textsf{DESC} before (denoted ``circ'' with yellow solid lines) and after (denoted ``opt'' with purple solid lines) the  optimization. Also shown as red dashed lines are the flux surfaces from the local Miller model \eqref{eq:theory_miller}, where the relevant parameters are listed in table~\ref{table:miller}. Lengths are normalized by $R_0$. (c,d) Comparison of \textsf{cvdrift} between the stellarator and \textsf{DESC} result at $r/a=0.5$ before and after the optimization. (e,f) Comparison of \textsf{cvdrift} at $r/a=0.5$ between the stellarator and the Miller model.}
    \label{fig:equiv_LBD_QS}
\end{figure}

Next, we find equivalent tokamaks for the $\beta=2.5\%$ QA and QH configurations reported in \cite{LBD22}. Compared to the precise QS configurations, these configurations are less accurately described by the NAE. In particular, the $\iota$ profile is non-monotonic, so that $\hat{s}$ is not small and varies significantly at different $r$. Nevertheless, other geometric quantities such as $B$ and $p$ can still be approximately described from the NAE, so that we can obtain the coefficients $p_2$, $B_{2\rmc}$, and $B_{20}$ through linear fitting in $r^2$. The relevant parameters are described in the caption of figure~\ref{fig:LBD_QS_B20}. To the zeroth order, the equivalent tokamaks are circular and
are characterized by
\begin{align}
\label{eq:application_LBD_QA}
    &{\rm QA}: \bigg|\frac{R_0^\rmt}{a^\rmt}\bigg|=9.39,\quad
    \iota^\rmt=-0.59,\quad \hat{s}^\rmt=-0.45, 
    \\
\label{eq:application_LBD_QH}
    &{\rm QH}: \bigg|\frac{R_0^\rmt}{a^\rmt}\bigg|=5.03,\quad \iota^\rmt=-1.62, \quad
    \hat{s}^\rmt=-0.028.    
\end{align}
Here, we use $\iota^\rms$ and $\hat{s}^\rms$ at $r/a=0.5$ as their representative values. Similar to the precise QS configurations in section~\ref{sec:application_precise}, here the equivalent tokamak for the QH is more compact than the QA due to its larger $|\bar{\eta}^\rms|$, and has a larger $|\iota^\rmt|$. Meanwhile, the equivalent tokamak for the QA has a negative $\hat{s}^\rmt$, while for the QH $\hat{s}^\rmt$ remains small. The global MHD equilibria for the equivalent tokamaks are generated from \textsf{DESC} with fixed shape and toroidal flux at the outermost flux surface, fixed $\iota$ profile, and fixed $p$ profile. Here, the outermost flux surfaces are circular with radius $a^\rmt$,  and the $\iota$ profiles are the corresponding stellarator $\iota_N=\iota -N$ profiles divided by $\bar{\eta}^\rms R_0^\rms$. Also, even though the pressure gradient enters as a first-order term, we include the finite $p$ here, which is the same as the corresponding stellarator so that the difference between \textsf{gbdrift} and \textsf{cvdrift} can be accounted for. The results are plotted in figures~\ref{fig:equiv_LBD_QS}(a,b). Due to the Shafranov shift, the flux surfaces are not strictly (but still close to) circular. Comparisons of \textsf{cvdrift} between these circular equivalent tokamaks and their corresponding stellarators are shown in  figures~\ref{fig:equiv_LBD_QS}(c,d), and it is seen that because $\hat{s}$ matches between the tokamak and the stellarator, the secular behavior in $\vartheta$ can be correctly described.

To find equivalent tokamaks with first-order accuracy, the relation \eqref{eq:NAE_B20_match} must be satisfied, and the results are shown in figure~\ref{fig:LBD_QS_B20}. The solutions are $\bar{\eta}^\rmt=-0.47$ for the QA and $\bar{\eta}^\rmt=-0.37$ or $-1.08$ for the QH. However, the corresponding global MHD equilibria generated from \textsf{DESC} do not provide satisfactory agreement in the geometric coefficients, mainly for two reasons. First, $\hat{s}$ is negative and large for the QA configuration, while $\hat{s}$ remains positive and small for the NAE configuration. Second, for the QH, the global equilibria quickly deviate from the NAE descriptions as we move away from the axes. Therefore, we no longer attempt to generate the equilibria based on the NAE, but rather aim at improving the agreement between the equivalent tokamaks and the stellarators as an optimization problem. An objective function has been implemented in \textsf{DESC}, which minimizes the differences in geometric quantities between the tokamak and the stellarator \footnote{\url{https://github.com/PlasmaControl/DESC/tree/rg/HX_geometry}}. Then, the optimization is carried out by varying the coefficients of the boundary shape and solving the MHD force balance equation, while fixing the $\iota$ profile, the $p$ profile, $B_0$, and $\psi_{\rm edge}$. Here, we focus on \textsf{cvdrift} as the objective function. The flux-surface shapes after the optimization are shown in figures~\ref{fig:equiv_LBD_QS}(a,b), and as seen in figures~\ref{fig:equiv_LBD_QS}(c,d)  the agreement is improved after the optimization. We note that as the flux surfaces deviate from circular,  $\bar{\eta}^\rmt$ will no longer equal $1/R_0^\rmt$, and hence $a^\rmt$ and $\iota^\rmt$ should be adjusted accordingly as in \eqref{eq:NAE_zeroth_order}. More self-consistent global optimizations will be left as future work.

In addition, we also seek a local Miller equilibrium that provides better agreement in \textsf{cvdrift} between the stellarator and the tokamak. This is done as a bounded nonlinear least-squares optimization problem using the trust-region reflective  algorithm implemented in \textsf{least\_squares} from \textsf{SciPy}. During the optimization we fix $\rho=0.5$ and $\beta'$ (as the difference between \textsf{gbdrift} and \textsf{cvdrift} from the stellarator), while varying $R_0$, $R_0'$, $q$, $\hat{s}$, $\varkappa$, $\varkappa'$, $\Delta$, and $\Delta'$. The optimization minimizes the weighted residuals between the stellarator and the Miller geometry for the geometric quantities listed  from \eqref{eq:NAE_bmag_gradpar} to \eqref{eq:NAE_cvdrift}. The optimized Miller parameters are listed in table~\ref{table:miller}, and the corresponding flux-surface  shapes are shown in figures~\ref{fig:equiv_LBD_QS}(a,b), which agree with the \textsf{DESC} results after optimization, indicating that the optimizer from \textsf{DESC} was finding the equivalent tokamaks in the correct direction. Also, as shown in figures~\ref{fig:equiv_LBD_QS}(e,f), the agreement in \textsf{cvdrift} is better, and we note that the same level of agreement is achieved for the other geometric quantities as well for the QA; for the QH, however, we found it harder to achieve the same level of good agreement for \textsf{cvdrift0} without compromising other quantities, so its weight was set to a lower value.

The above results demonstrate that one can find an equivalent tokamak either as a global equilibrium or as a local Miller equilibrium, and the latter can provide better agreement locally. Also note that the same Miller optimization has been applied to the precise QA configuration but no parameters were found that achieve good agreement in \textsf{cvdrift}. This is consistent with the NAE result in section~\ref{sec:application_precise}  that no solution could be found from the NAE theory for the precise QA configuration.

\section{Numerical results from global and local gyrokinetic simulations}
\label{sec:numerical}
\subsection{Equivalent tokamaks for the $\beta=2.5\%$ QA and QH configurations}
\label{sec:numerical_LBD}
Here, we simulate the ITG turbulent transport of  the $\beta=2.5\%$ QA and QH stellarators and their equivalent tokamaks using the global gyrokinetic code \textsf{GTC}. For simplicity, we use the zeroth-order equivalent tokamaks with circular flux surfaces as shown in figures~\ref{fig:equiv_LBD_QS}(a,b), and they both have $B_0^\rmt=5.7~{\rm T}$ and $a^\rmt=1.7~{\rm m}$, consistent with the stellarators. The simulation setup is similar but not identical to that reported in \cite{Chen25}. We use deuterium ions, whose on-axis density and temperature are $n_0=10^{19}~{\rm m}^{-3}$ and $T_0=50~{\rm keV}$, respectively;  electrons are assumed to be adiabatic and have the same density and temperature as ions. Define the local density and temperature gradient scale lengths as $L_n =(-rB_0\pd_\psi\ln n_\rmi)^{-1}$ and $L_T= (-rB_0\pd_\psi\ln T_\rmi)^{-1}$, and refer to $a/L_n$ and $a/L_T$ as the local gradients. The ions have zero density gradient, $a/L_n=0$, while their temperature gradient has the following form:
\begin{equation}
    \frac{a}{L_T(r)}=\frac{1}{2}\frac{a}{L_{T0}}\left(\tanh\frac{r-0.3a}{\Delta r}+\tanh\frac{0.7a-r}{\Delta r}\right),
\end{equation}
where $\Delta r=0.02a$, and we choose $a/L_{T0}=2$. The simulation domain is $r/a=[0.1,0.9]$ in the radial direction, and $\zeta\in[0,2\pi/N_{\rm fp})$ in the toroidal direction, with $N_{\rm fp}=2$ for the QA and $N_{\rm fp}=4$ for the QH. The radial, poloidal, and toroidal resolutions are 200, 2000, 27, respectively, which suffice to give converged results on the radial heat fluxes. Collisions are not included in the simulations.

\begin{figure}
    \centering
    \includegraphics[width=1\linewidth]{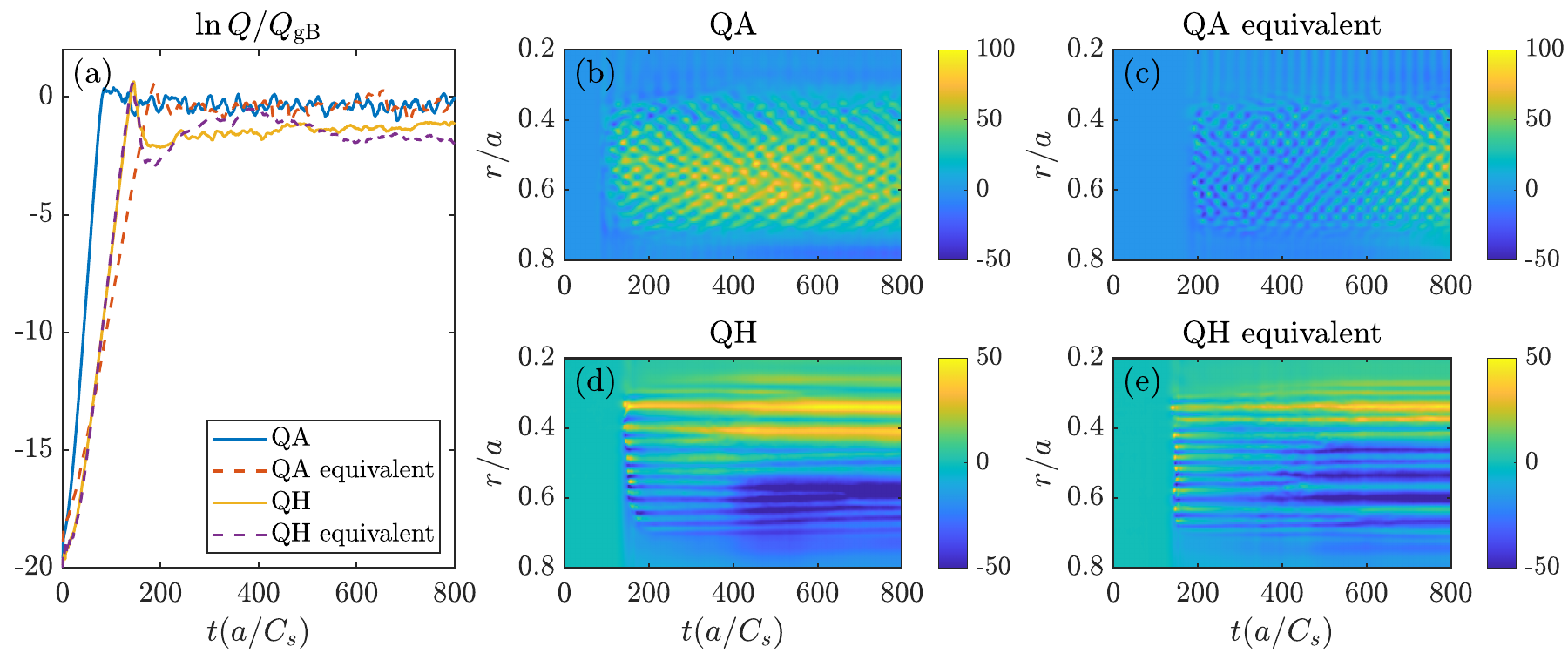}
    \caption{Comparison of ITG turbulent transport of the $\beta=2.5\%$ QA and QH configurations and their equivalent tokamaks from global \textsf{GTC} simulations. (a) The radial heat fluxes (logarithmic plot) in units of $Q_{\rm gB}$ versus time across the $r/a=0.5$ flux surfaces. (b-e) The zonal radial electric fields \eqref{eq:numerical_zonal_Er} in units of kV/m versus radius and time.}
    \label{fig:LBD_equiv_simulation}
\end{figure}

The radial heat fluxes in gyro-Bohm units $Q_{\rm gB}=(n_0v_\rmt T_0)(\rho_\rmi/a)^2$ are plotted in figures~\ref{fig:LBD_equiv_simulation} ($v_\rmt=\sqrt{T_0/m_\rmi}$ and $\rho_\rmi=m_\rmi v_\rmt/eB_0$), which grow linearly and then nonlinearly saturate. The heat flux of the QA configuration is larger than that of the QH configuration. \cite{Chen25} argued that this difference can be explained from the zonal flows, which are oscillatory in the QA but stationary in the QH. This observation is verified from figures~\ref{fig:LBD_equiv_simulation}(b,d), where we plot the zonal electric field defined as
\begin{equation}
    E_r=-rB_0\pd_\psi\avg{\Phi},\label{eq:numerical_zonal_Er}
\end{equation}
where $\Phi$ is the perturbation in the electrostatic potential and $\avg{\dots}$ denotes the flux-surface average. The magnitudes of $E_r$ are similar between the QA and the QH, but $E_r$ in the QA oscillates in radius and time, reminiscent of the recently discovered ``toroidal secondary mode''  \citep{Nies26a,Nies26b}. In contrast, $E_r$ in the QH is stationary and hence expected to be more effective at suppressing the turbulence. Also, as seen from figure~\ref{fig:LBD_equiv_simulation}(c,e), the behaviors of $E_r$ are the same in the equivalent tokamaks.

We briefly discuss the possible physical mechanism of zonal flows in the following, focusing on the role of $\iota$ and $\hat{s}$. The Rosenbluth--Hinton (RH) residual flow is considered a good indicator for the ITG turbulent transport \citep{Rosenbluth98,Plunk24,Zhu25,Chen25,Wei26}. The RH level in QS stellarators can be calculated as 
\begin{equation}
    \frac{E_r(t=\infty)}{E_r(t=0)}\bigg|_{\rm stellarator}=\frac{1}{1+1.6/\left[(\epsilon^\rms)^{1/2}(\iota_N^\rms)^2\mc{C}\right]},
\end{equation}
where $\epsilon^\rms=|\bar{\eta}^\rms r^\rms|$  and $\mc{C}=\avg{|\nabla r|^2}/(\bar{\eta}^\rms R_0^\rms)^2$ \citep{Zhu25}. Using the relation $\epsilon^\rmt=\epsilon^\rms$ and $\iota^\rmt=\iota_N^\rms/(\bar{\eta}^\rms R_0^\rms)$ for the circular equivalent tokamaks, it can be rewritten as
\begin{equation}
    \frac{E_r(t=\infty)}{E_r(t=0)}\bigg|_{\rm stellarator}=\frac{1}{1+1.6/\left[(\epsilon^\rmt)^{1/2}(\iota^\rmt)^2\avg{|\nabla r|^2}\right]}.
\end{equation}
In comparison, the RH level for the circular tokamaks is
\begin{equation}
        \frac{E_r(t=\infty)}{E_r(t=0)}\bigg|_{\rm tokamak}=\frac{1}{1+1.6/\left[(\epsilon^\rmt)^{1/2}(\iota^\rmt)^2\right]}.
\end{equation}
Therefore, the two RH levels differ by a geometric factor $\avg{|\nabla r|^2}$, whose value is stellarator-specific and not captured by the equivalent tokamaks. Nevertheless, the RH level depends most strongly on $\iota$, and this dependency is correctly captured. In particular, compared to the QA, the QH and its equivalent tokamak have much larger $|\iota_N|$ and hence a much larger RH level. The oscillatory zonal flows observed for the QA are similar to the toroidal secondary mode discussed by \cite{Nies26a,Nies26b}, who also mentioned that this mode only exists at $|\iota|<1$. The reason is that the toroidal secondary mode arises due to an up-down asymmetric pressure perturbation associated with the radial heat transport from the ITG; at $|\iota|>1$, field lines are ``short circuited'' due to the small parallel connection length $|R/\iota|$, and hence the pressure perturbations are quickly neutralized by the parallel heat fluxes. Similar behaviors have been observed for the geodesic-acoustic mode oscillations in QS stellarators, which disappear at $|\iota|>1$ \citep{Zhu25}. Therefore, for the QH and its equivalent tokamak, zonal flows are stationary due to the lack of an oscillation mechanism. Finally, the QA and its equivalent tokamak have negative $\hat{s}$, which is consistent with another recent study of ITG turbulent transport by \cite{Yang24}, who reported that zonal flows are weaker and oscillatory at negative $\hat{s}$ compared to those at positive $\hat{s}$.
\subsection{Equivalent tokamaks for a QH database}
\begin{figure}
    \centering
    \includegraphics[width=1\linewidth]{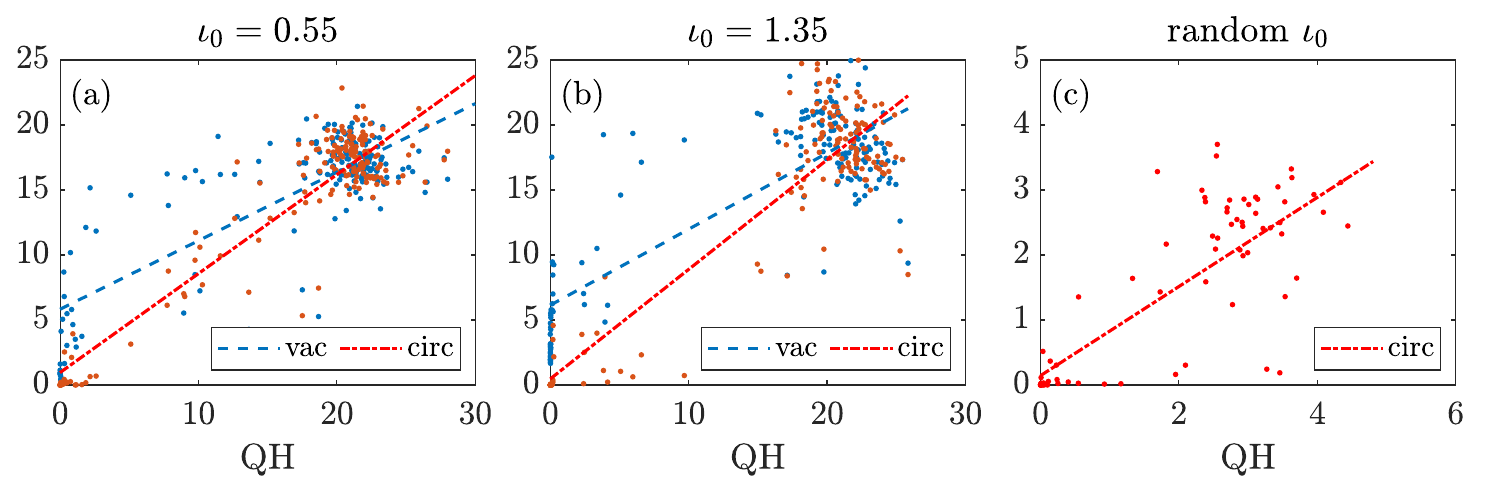}
    \caption{Comparison of ITG turbulent transport of QH configurations and their equivalent tokamaks from a database \citep{Wei26}.  (a,b): \textsf{GX} simulation results for the $\iota_0=0.55$ and $\iota_0=1.35$ configurations. (c): GTC simulation results for the random-$\iota_0$ configurations. The horizontal axes show the heat fluxes $Q/(Q_{\rm gB}\avg{|\nabla r|})$ from the QH configurations, while the vertical axes show the heat fluxes from the equivalent tokamaks with zero pressure (denoted ``vac'', blue dots) or with the same pressure profile as the stellarator (denoted ``circ'', red dots). The blue dashed lines and the red dashed-dotted lines are the linear fits, and for the ``circ'' cases the fitting formulas are (a) $y=0.76x+0.95$, (b) $y=0.84 x+0.46$, and (c) $y=0.69x+0.14$.}
    \label{fig:QH_database_simulation}
\end{figure}

Here, we simulate the ITG turbulent transport in QH configurations and their equivalent tokamaks from a recently created database \citep{Wei26}. This database is generated using \textsf{DESC}, where each configuration is initialized with a random axis shape, a random $p$ profile, and either a random or a fixed $\iota$ profile; then, the configuration is optimized until good QS is achieved. For the configurations considered below, both the two-point QS error and the normalized force-balance error are smaller than 10\%. We first use the local gyrokinetic code \textsf{GX} to simulate two groups of configurations, whose $\iota_0$ are fixed at $0.55$ and $1.35$, respectively. For each configuration, we generate its zeroth-order circular equivalent tokamaks from \textsf{DESC} using the same procedure described in section~\ref{subsec:application_LBD}. Namely, $a^\rmt$ and the $\iota^\rmt$ profile are calculated from \eqref{eq:NAE_zeroth_order_circ}, while the $p^\rmt$ profile is chosen to be either zero (vacuum) or the same as the stellarator. Then, the tokamak equilibrium is solved with a fixed circular shape and fixed toroidal flux at the boundary. The \textsf{GX} simulations are carried out in flux tubes centered at $r/a=0.5$, whose sizes in the radial and bi-normal directions are $L_x\approx L_y=30\pi\rho_\rmi$; the length in the parallel direction is $\theta\in(-2\pi,2\pi)$, where the periodic boundary condition is applied at the ends for simplicity. The grid resolution is $(96,96,48)$ in $(x,y,\theta)$ and  $(16,8)$ in $(v_\parallel,\mu)$. Ions have a temperature gradient $a/L_T=4.7$, density gradient $a/L_n=0$, and collision rate $\nu_{\rmi\rmi}=0.01 v_\rmt/a$, while electrons are adiabatic. A total of 350 configurations are simulated with $\iota_0=0.55$, and we pick 180 of them where the ITG  is linearly unstable and the nonlinear heat fluxes reach steady levels; for the $\iota_0=1.35$ configurations, we pick 166 simulations out of 300 under the same criteria. The time-averaged heat fluxes are plotted in figures~\ref{fig:QH_database_simulation}(a,b), which show good linear correlation between the original QH configurations and the equivalent tokamaks. The Pearson correlation coefficients are around 0.81 between the QH and the tokamaks with $p^\rmt=0$, and increase to around 0.92 between the QH and the tokamaks with $p^\rmt\neq 0$. Therefore, even though $p^\rmt$ does not enter the zeroth-order NAE description, its inclusion will improve the correlation, which is also visually verified from the linear fits in the figures.

We note that the heat fluxes outputted by \textsf{GX} and plotted in figure~\ref{fig:QH_database_simulation} are $Q/(Q_{\rm gB}\avg{|\nabla r|})$ instead of $Q/Q_{\rm gB}$, where the extra geometric factor $\avg{|\nabla r|}$ is around 1.4 for the QH configurations but around 1.0 for the circular tokamaks.  The factor $\avg{|\nabla r|}$ is stellarator-specific and measures the level of flux-surface compression, and is considered to be the most important factor that determines the heat fluxes \citep{Landreman25}. Our results here also indicate that the heat fluxes from the QH stellarators can be estimated as the heat fluxes of their equivalent tokamaks multiplied by a factor $\avg{|\nabla r|}\approx 1.4$.

As a comparison, we also performed global gyrokinetic \textsf{GTC} simulations on another group of QH stellarators with random $\iota_0$ in the database. A total of 220 configurations are simulated, which on average have $R_0=1~{\rm m}$, $a=0.3~{\rm m}$, and $B_0=0.2~{\rm T}$. We simulate hydrogen ions with on-axis density $n_0=10^{19}~{\rm m}^{-3}$ and temperature $T_0=5~{\rm eV}$; the latter is small due to the small $a$ and $B_0$. The ions have a temperature gradient $a/L_T=4.7$ and density gradient $a/L_n=0$, which are consistent with the \textsf{GX} simulations, but collisions are not included; electrons are also assumed adiabatic. The simulation domain is $r/a=[0.2,0.8]$ in the radial direction and $\zeta\in[0,2\pi/N_{\rm fp})$ in the toroidal direction with $N_{\rm fp}=4$. The radial, poloidal, and toroidal resolutions are 150, 1000, 9, respectively, which are relatively low due to the constraints on computational resources. We pick 84 well-behaved cases and the results are shown in figure~\ref{fig:QH_database_simulation}(c), where the heat fluxes are also divided by $\avg{|\nabla r|}$ to be consistent with other figures. Similar to the \textsf{GX} results, good agreement is found in the heat fluxes between QH stellarators and their equivalent tokamaks with a Pearson coefficient of 0.81. But we also found that the heat fluxes from \textsf{GTC} are lower than those from \textsf{GX}, either because of the different configurations used or the discrepancy between local and global simulations. Due to the low resolution for these global simulations, we leave possible explanations for future work.
\section{Conclusions}
\label{sec:conclusion}
In this work, a systematic theory of the equivalent tokamaks for QS stellarators is presented based on the local gyrokinetic formulation and the NAE theory. It is shown that to zeroth-order accuracy in the minor radius the equivalent tokamaks can be chosen to have circular flux surfaces and can be characterized by three geometric quantities: the aspect ratio $R/a$, the rotational transform $\iota$, and the magnetic shear $\hat{s}$. To achieve first-order accuracy, however, not all QS stellarators have equivalent tokamaks, but good approximations can be found for some cases either as global equilibria generated from \textsf{DESC} or as local equilibria described by the Miller model. Simulations of ITG turbulent transport are performed for a selection of QS configurations using the local gyrokinetic code \textsf{GX} and global gyrokinetic code \textsf{GTC}, which demonstrate quantitative agreement between the stellarators and their equivalent tokamaks. 

\section*{Acknowledgments}
H.Z. thanks H. Chen, W. Sengupta, A. Brown, and  D. Kennedy  for useful discussions.  The artificial intelligence model ChatGPT was used to assist with equation derivations and manuscript proofreading.
\section*{Data availability statement}
The data that supports the findings of this study are openly available at Zenodo \citep{Zhu26data}.

\section*{Declaration of interests}
The authors report no conflict of interest.
\appendix
\section{Derivation of $B_{20}$ for tokamaks from the near-axis expansion}
\label{app:B20}
Here, we derive $B_{20}$ for tokamaks following Appendix A of \cite{Landreman19b}. Due to axisymmetry, things are greatly simplified since many geometric coefficients involving toroidal derivatives become zero. We will also assume $B_0=R_0=1$ to further simplify the coefficients, which also leads to $G_0=B_0R_0=1$ and $\kappa=R_0^{-1}=1$. At first order, $\iota_0$ is obtained from equation~\eqref{eq:NAE_sigma_tokamak}:
\begin{equation}
    \iota_0=\frac{2G_0I_2\bar{\eta}^2}{1+\bar{\eta}^4}.
\end{equation}
Other coefficients are $B_{1\rmc}=\bar{\eta}$, $B_{1\rms}=0$, $X_{1\rmc}=\bar{\eta}$, $Y_{1\rmc}=0$, $X_{1\rms}=0$, $Y_{1\rms}=\bar{\eta}^{-1}$, and $\delta_0=\delta_{1\rmc}=0$. The three coefficients from equations~(A24,A30,A31) of \cite{Landreman19b} are
\begin{equation}
    V_1=X_{1\rmc}^2+Y_{1\rms}^2=\bar{\eta}^2+\bar{\eta}^{-2}, \quad V_2=0,\quad V_3=X_{1\rmc}^2-Y_{1\rms}^2=\bar{\eta}^2-\bar{\eta}^{-2}.
\end{equation}
At second order, we have two additional free parameters $p_2$ and $B_{2\rmc}$ in the MHD force balance \eqref{eq:NAE_MHD}. Other coefficients are $Z_{20}=Z_{2\rmc}=0$, $Z_{2\rms}=\iota_0 V_3/4$, and $B_{2\rms}=0$. The four coefficients from equations~(A37-A40) of \cite{Landreman19b} are
\begin{equation}
    q_\rms=-\iota_0X_{1\rmc}=-\iota_0\bar{\eta},\quad q_\rmc=r_\rms=0,\quad r_\rmc=\iota_0Y_{1\rms}=\iota_0\bar{\eta}^{-1}.
\end{equation}
Then, from equations~(A35,A36,A32,A33) of \cite{Landreman19b}, we have the following system of equations
\begin{align}
&X_{2\rms}=0,\\
\notag
&X_{2\rmc}= 2\iota_0Z_{2\rms}-\left(-B_{2\rmc}+\frac{3B_{1\rmc}^2}{4}-\frac{X_{1\rmc}^2}{4}-\frac{-q_\rms^2+r_\rmc^2}{4}\right)
\\
\notag
&\qquad =\frac{\iota_0^2V_3}{2}+B_{2\rmc}-\frac{3}{4}\bar{\eta}^2+\frac{1}{4}\bar{\eta}^2-\frac{1}{4}\iota_0^2\bar{\eta}^2+\frac{1}{4}\iota_0^2\bar{\eta}^{-2}
\\
&\qquad =\frac{1}{4}\iota_0^2(\bar{\eta}^2-\bar{\eta}^{-2})-\frac{1}{2}\bar{\eta}^2+B_{2\rmc},
\\
&X_{1\rmc}Y_{2\rmc}-X_{1\rmc}Y_{20}=0,
\\
&-\frac{1}{2}X_{1\rmc}=X_{1\rmc}Y_{2\rms}+X_{2\rmc}Y_{1\rms}+X_{20}Y_{1\rms},
\end{align}
and the last equation gives 
\begin{equation}
X_{2\rmc}+X_{20}=-X_{1\rmc}\left(\frac{1}{2}+Y_{2\rms}\right)/Y_{1\rms}=-\left(\frac{1}{2}+Y_{2\rms}\right)\bar{\eta}^2.
\end{equation}
From equations~(A41,A42) of \cite{Landreman19b}, it turns out that $Y_{20}=Y_{2\rmc}=0$, and
\begin{align}
    X_{1\rmc}f_{X\rms}-Y_{1\rms}f_{Y0}-Y_{1\rms}f_{Y\rmc}=0,\\
    -X_{1\rmc}f_{X0}+X_{1\rmc}f_{X\rmc}+Y_{1\rms}f_{Y\rms}=0,
\end{align}
so that
\begin{equation}
    f_{Y0}+f_{Y\rmc}=f_{X\rms}X_{1\rmc}/Y_{1\rms}=\bar{\eta}^2 f_{X\rms},
\end{equation}
with the coefficients
\begin{align}
    &f_{X0}=f_{X\rmc}=f_{Y\rms}=0,
    \\
    &f_{X\rms}=-2\iota_0X_{2\rmc}+Z_{2\rms}-I_2\left(\frac{1}{2}X_{1\rmc}Y_{1\rms}-2Y_{2\rms}\right)-\frac{1}{2}\delta_{1\rms}Y_{1\rms},
    \\
    &f_{Y0}=4X_{2\rmc}Z_{2\rms}+I_2\left(\frac{1}{2}X_{1\rmc}^2-2X_{20}\right)-\frac{1}{2}\delta_{1\rms}X_{1\rmc},
    \\
    &f_{Y\rmc}=2\iota_0 Y_{2\rms}+4X_{20}Z_{2\rms}+I_2\left(\frac{1}{2}X_{1\rmc}^2-2X_{2\rmc}\right)+\frac{1}{2}\delta_{1\rms}X_{1\rmc}.
\end{align}
The above algebraic equations result in
\begin{equation}
    Y_{2\rms}=\left[\bar{\eta}^2\left(-2\iota_0X_{2\rmc}+3Z_{2\rms}-\frac{5}{2}I_2\right)-\frac{1}{2}\delta_{1\rms}\bar{\eta}\right]\bigg/\left(2\iota_0-4Z_{2\rms}\bar{\eta}^2\right).
\end{equation}
Finally, from (A34) of
\cite{Landreman19b}:
\begin{align}
    X_{20}&=-\left(-B_{20}+\frac{3B_{1\rmc}^2}{4}+\left(G_2+\iota_0 I_2\right)-\frac{X_{1\rmc}^2}{4}-\frac{q_\rms^2+r_\rmc^2}{4}\right)\notag
    \\
    &=B_{20}-\frac{3}{4}\bar{\eta}^2+\mu_0 p_2+\frac{1}{4}\bar{\eta}^2+\frac{1}{4}\iota_0^2\left(\bar{\eta}^2+\bar{\eta}^{-2}\right).
\end{align}
Writing $B_{20}$ in terms of $X_{20}$ and the other coefficients, we reach the final expression \eqref{eq:NAE_B20}.    

\bibliographystyle{jpp}

\bibliography{refs}

\end{document}